\documentclass[aps,pre,10pt,amsmath,amssymb,floatfix]{revtex4-2}

\usepackage{graphicx}
\usepackage{booktabs}
\usepackage{braket}
\usepackage{hyperref}

\hypersetup{
    colorlinks=true,
    linkcolor=black,
    citecolor=black,
    urlcolor=black,
    breaklinks=true,
    pdfborder={0 0 0},
    pdfusetitle=true
}
\allowdisplaybreaks

\begin{document}

\title{Quantum Otto engine with field-decoupled idle levels in a non-Hermitian XY model}

\author{Maimaitiyiming Tusun}
\email{hawk@xjnu.edu.cn}
\author{Fang Zhao}
\affiliation{School of Physics and Electronic Engineering, Xinjiang Normal University, Xinjiang Key Laboratory of Luminescent Minerals and Optical Functional Materials, Urumqi 830054, China}

\date{\today}

\begin{abstract}
We propose a microscopic realization of the idle-level quantum heat engine architecture based on a two-qubit non-Hermitian XY model with a staggered imaginary magnetic field. The energy spectrum naturally separates into two working levels coupled to the external field and two field-decoupled idle levels entirely independent of it. Tuning the non-Hermitian parameter drives the system from a dissipative accelerator regime into a genuine heat engine regime, while simultaneously enhancing both the net work output and the cycle efficiency. The efficiency enhancement originates from the compression of the idle-level gap, which redistributes level occupations and progressively suppresses the idle-level heat current, signalling the approach to the reverse heat-flow regime that underlies the idle-level engine principle. We map the full thermodynamic phase diagram identifying engine, refrigerator, accelerator, and heater regimes, and decompose the heat current into working-level and idle-level contributions to clarify the microscopic origin of the performance gain. The model is implementable in trapped-ion and NMR quantum simulators. These results establish non-Hermiticity as a versatile control knob for idle-level quantum thermal machines.
\end{abstract}


\maketitle

\section{Introduction}
\label{sec:intro}

Quantum thermodynamics investigates energy conversion at the nanoscale and has established firm theoretical foundations over the past decades~\cite{Alicki1979,Kosloff1984,Geva1992}. Quantum heat engines have been experimentally realized in trapped-ion systems~\cite{Abah2012,Rossnagel2016}, nitrogen-vacancy centers~\cite{Klatzow2019}, and nuclear magnetic resonance setups~\cite{Peterson2019}, confirming many predictions of the quantum Otto cycle paradigm~\cite{Quan2007,Kosloff2017}. Comprehensive overviews can be found in textbooks and review articles~\cite{Gemmer2009,Vinjanampathy2016,Goold2016,Binder2018}.

In parallel, non-Hermitian quantum systems with parity-time (PT) symmetry have attracted broad interest owing to their real spectra and unconventional phenomenology~\cite{Bender1998,Bender2007}. Applications range from photonics~\cite{Feng2017,Ozdemir2019,Ruter2010,Konotop2016,ElGanainy2018} to ultracold atomic gases~\cite{Ashida2020}. Extending thermodynamics to non-Hermitian settings has revealed effects such as enhanced output power and modified efficiency bounds~\cite{Lin2016,Insinga2018,Zhang2022}, as well as distinctive work statistics near exceptional points~\cite{Zhou2021,Khandelwal2021}.

A conceptually distinct direction is the idle-level heat engine proposed by de Oliveira and Jonathan~\cite{DeOliveira2021}. In this architecture, a subset of energy levels remains decoupled from the external work parameter and acts as an internal entropy reservoir. By adjusting the idle-level gap, the efficiency can be tuned continuously from zero up to the standard Otto bound, and the cycle can be switched between engine, refrigerator, accelerator, and heater regimes. The underlying mechanism is a reverse heat flow through the idle levels that effectively recycles energy between the two thermal baths. Fluctuation properties of this class of engines were studied in subsequent work~\cite{DeAlmeida2024}. The nonadiabatic regime of idle-level engines was investigated in Ref.~\cite{Cherubim2022}, where it was shown that qualitative features are robust against finite-time effects.

Although the idle-level principle is theoretically well established, a concrete microscopic realization within a realistic spin Hamiltonian has been lacking. The original proposal employed a coupled-spin model where the exchange coupling serves as the control knob, but tuning the coupling simultaneously modifies both the idle and working levels, so the control is not fully independent. A model in which the idle levels are strictly decoupled from the work parameter and controlled by a separate knob would be both conceptually cleaner and experimentally more versatile.

The non-Hermitian two-qubit XY model with a staggered imaginary field provides precisely such a realization. Its entanglement properties were studied by Li et al.~\cite{Li2023}, and its spectrum splits algebraically into a field-dependent working pair and a field-decoupled idle pair controlled solely by the non-Hermitian parameter. This structure matches the idle-level architecture perfectly, but its thermodynamic and heat-engine properties have not been explored.

In this paper we construct a quantum Otto cycle based on this model. Working within the PT-unbroken phase and the biorthogonal canonical thermodynamic framework~\cite{Gardas2016}, we derive analytical expressions for all cycle quantities. We show that varying the non-Hermitian parameter switches the cycle from a dissipative accelerator regime to a heat engine regime and significantly enhances the efficiency. We map the full thermodynamic phase diagram, decompose the heat current into working-level and idle-level contributions, and explicitly connect the microscopic occupation redistribution to the macroscopic reverse heat-flow mechanism. We also discuss experimental implementation on trapped-ion platforms and address the roles of postselection cost and finite-time nonadiabatic losses.

\section{The model}
\label{sec:model}

We consider the same Hamiltonian as in Ref.~\cite{Li2023},
\begin{equation}
H=-\frac{J}{2}\bigl[(1+\gamma)(\sigma_1^x\sigma_2^x+\sigma_2^x\sigma_1^x)
+(1-\gamma)(\sigma_1^y\sigma_2^y+\sigma_2^y\sigma_1^y)\bigr]
-h(\sigma_1^z+\sigma_2^z)+i\eta(-\sigma_1^z+\sigma_2^z),
\label{eq:H1}
\end{equation}
where $J>0$ sets the energy scale, $\gamma$ is the exchange anisotropy, $h$ is the real transverse field, and $\eta$ is the strength of the staggered imaginary field. Such non-Hermitian terms can be engineered in photonic waveguide arrays~\cite{Feng2017} and circuit quantum simulators~\cite{Ozdemir2019} via balanced gain and loss.

Defining dimensionless variables $h_0=h/J$ and $\eta_0=\eta/J$, the Hamiltonian reads
\begin{equation}
H=-J\bigl[(1+\gamma)\sigma_1^x\sigma_2^x+(1-\gamma)\sigma_1^y\sigma_2^y\bigr]
-Jh_0(\sigma_1^z+\sigma_2^z)+iJ\eta_0(-\sigma_1^z+\sigma_2^z).
\label{eq:H2}
\end{equation}

In the computational basis $\{\ket{\uparrow\uparrow},\ket{\uparrow\downarrow},
\ket{\downarrow\uparrow},\ket{\downarrow\downarrow}\}$, $H$ is block diagonal,
\begin{equation}
H=\begin{pmatrix}
-2Jh_0 & 0 & 0 & -2J\gamma\\
0 & -2iJ\eta_0 & -2J & 0\\
0 & -2J & 2iJ\eta_0 & 0\\
-2J\gamma & 0 & 0 & 2Jh_0
\end{pmatrix}.
\label{eq:Hmat}
\end{equation}

The $\{\ket{\uparrow\uparrow},\ket{\downarrow\downarrow}\}$ block yields the working-level eigenvalues
\begin{equation}
E_{1,2}=\mp 2J\sqrt{h_0^2+\gamma^2},
\label{eq:Ework}
\end{equation}
while the $\{\ket{\uparrow\downarrow},\ket{\downarrow\uparrow}\}$ block yields the idle-level eigenvalues
\begin{equation}
E_{3,4}=\mp 2J\sqrt{1-\eta_0^2}.
\label{eq:Eidle}
\end{equation}
Crucially, the idle eigenvalues depend only on $\eta_0$ and are completely independent of the external field $h_0$. The full biorthonormal eigenbasis is derived in Appendix~\ref{app:diag}.

For $|\eta_0|<1$ the entire spectrum is real and the Hamiltonian is PT symmetric~\cite{Mostafazadeh2010}. We restrict all calculations to this regime and further impose the bound $\eta_0<0.95$ so as to stay sufficiently far from the exceptional point at $\eta_0=1$, where the biorthogonal eigenbasis becomes ill-conditioned and the spectral decomposition becomes numerically unstable. This bound ensures that the biorthogonal normalization factor remains bounded and that the adiabatic condition for the idle subspace remains satisfied, as quantified in Appendix~\ref{app:adiabatic}.

\section{Thermodynamic framework}
\label{sec:thermo}

Following the standard non-Hermitian canonical thermodynamics formalism~\cite{Gardas2016}, the partition function for a system with a real discrete spectrum is
\begin{equation}
Z=\sum_{i=1}^4 e^{-\beta E_i},
\label{eq:Z}
\end{equation}
where $\beta=1/k_BT$ and we set $k_B=1$ for brevity. The Gibbs occupation probability of level $i$ is $p_i=e^{-\beta E_i}/Z$, and the internal energy is $U=\sum_i p_iE_i$. The identity $\langle H\rangle=U$ follows from the spectral decomposition $H=\sum_i E_i|\Psi_i^R\rangle\langle\Psi_i^L|$ and the biorthonormality condition $\langle\Psi_i^L|\Psi_j^R\rangle=\delta_{ij}$; details are given in Appendix~\ref{app:thermal}.

We emphasize that the biorthogonal Gibbs state does not describe ordinary thermalization with a Hermitian bath. It corresponds to an effective steady state of a continuously monitored and postselected ensemble, as routinely implemented in photonic and NMR simulators~\cite{Ashida2020}. All thermodynamic quantities in this work are defined for the conditioned ensemble of successful postselection events. In a realistic experiment, the postselection success probability decays over repeated cycles, which reduces the net work extractable per physical run; our quasistatic results represent the ideal conditional-ensemble limit, consistent with the standard framework in non-Hermitian quantum thermodynamics~\cite{Gardas2016}.

Introducing the level gaps
\begin{equation}
A(h)=2J\sqrt{h^2+\gamma^2},\qquad B(\eta_0)=2J\sqrt{1-\eta_0^2},
\label{eq:AB}
\end{equation}
the partition function simplifies to $Z=2\cosh(\beta A)+2\cosh(\beta B)$, and the internal energy becomes
\begin{equation}
U(h,T)=-\frac{A\sinh(\beta A)+B\sinh(\beta B)}{\cosh(\beta A)+\cosh(\beta B)}.
\label{eq:U}
\end{equation}
Here and throughout, $h$ denotes the dimensionless field $h/J$.

\section{The Otto cycle}
\label{sec:otto}

We consider a standard four-stroke quantum Otto cycle. Two idealizations are adopted: (i) the two adiabatic strokes are sufficiently slow that nonadiabatic transitions are negligible~\cite{Mostafazadeh2010}; (ii) the two isochoric strokes are sufficiently long for the system to relax to the biorthogonal Gibbs state at the corresponding bath temperature.

The four strokes proceed as follows.
\begin{enumerate}
\item Hot isochore: the system is held at $h=h_H$ and brought into contact with the hot bath at temperature $T_h$, reaching internal energy $U_1=U(h_H,T_h)$.
\item Adiabatic expansion: the field is ramped from $h_H$ to $h_C$ with the system isolated. Occupation probabilities are frozen, and the internal energy evolves to $U_2=\sum_i p_i^{(1)}E_i(h_C)$.
\item Cold isochore: the system is held at $h=h_C$ and brought into contact with the cold bath at temperature $T_c$, reaching internal energy $U_3=U(h_C,T_c)$.
\item Adiabatic compression: the field is ramped back from $h_C$ to $h_H$ with the system isolated. Occupation probabilities are frozen, and the internal energy returns to $U_4=\sum_i p_i^{(3)}E_i(h_H)$.
\end{enumerate}

The heat exchanges with the two baths and the net work per cycle are
\begin{equation}
Q_h=U_1-U_4,\qquad Q_c=U_2-U_3,\qquad W=Q_h-Q_c.
\label{eq:heat}
\end{equation}
When $W>0$ and $Q_h>0$, the cycle operates as a heat engine with efficiency $\eta=W/Q_h$. When $W<0$, $Q_h>0$, and $Q_c>0$, the cycle is in the accelerator (dissipative) regime, where external work and absorbed heat are both rejected to the cold bath~\cite{DeOliveira2021}.

After algebraic simplification (see Appendix~\ref{app:workheat}), the net work takes the form
\begin{equation}
W=(A_H-A_C)\left[
\frac{\sinh(\beta_c A_C)}{\cosh(\beta_c A_C)+\cosh(\beta_c B)}
-\frac{\sinh(\beta_h A_H)}{\cosh(\beta_h A_H)+\cosh(\beta_h B)}
\right],
\label{eq:W}
\end{equation}
and the absorbed heat is
\begin{equation}
Q_h=\frac{A_H\sinh(\beta_c A_C)+B\sinh(\beta_c B)}
{\cosh(\beta_c A_C)+\cosh(\beta_c B)}
-\frac{A_H\sinh(\beta_h A_H)+B\sinh(\beta_h B)}
{\cosh(\beta_h A_H)+\cosh(\beta_h B)}.
\label{eq:Qh}
\end{equation}

A key observation is that the idle gap $B$ appears only in the denominators of the net work expression, whereas it appears in both numerator and denominator of the absorbed heat. This asymmetry is the algebraic origin of the idle-level control mechanism: the idle levels modify the heat absorption much more strongly than they modify the net work, which allows efficiency to be tuned without altering the working-level energy difference $A_H-A_C$. In the Hermitian limit $\eta_0=0$, all expressions reduce smoothly to those of an ordinary four-level Otto engine.

\section{Results and discussion}
\label{sec:results}

For concreteness we fix the parameters to $J=1$, $\gamma=0.3$, $h_H=1.5$, $h_C=0.8$, $T_h=2.0$, and $T_c=0.5$, corresponding to a Carnot efficiency $\eta_C=1-T_c/T_h=0.75$.

\subsection{Mode switching and efficiency enhancement}

\begin{figure}[tb]
\centering
\includegraphics[width=0.95\columnwidth]{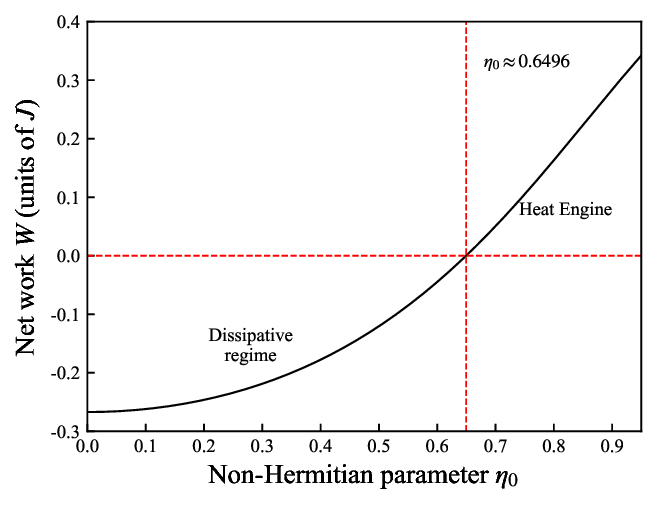}
\caption{Net work $W$ (in units of $J$) as a function of the non-Hermitian parameter $\eta_0$ (horizontal axis, range $0$ to $0.95$). The horizontal dashed line marks $W=0$ and the vertical line indicates the critical point $\eta_c\approx0.6496$. Parameters: $J=1$, $\gamma=0.3$, $h_H=1.5$, $h_C=0.8$, $T_h=2.0$, $T_c=0.5$.}
\label{fig1}
\end{figure}

Figure~\ref{fig1} shows the net work $W$ as a function of the non-Hermitian parameter $\eta_0$. For small $\eta_0$, $W$ is negative while both $Q_h$ and $Q_c$ remain positive, placing the cycle in the dissipative accelerator regime. As $\eta_0$ increases, the idle-level gap $B=2J\sqrt{1-\eta_0^2}$ shrinks, and the hyperbolic cosine terms in the denominators of Eq.~\eqref{eq:W} are suppressed, causing $W$ to rise monotonically. The net work vanishes at the critical value $\eta_c\approx0.6496$, above which the cycle operates as a genuine heat engine with positive work output.

The total entropy production per cycle, defined as the sum of the entropy changes of the two baths, is positive across the entire parameter range, consistent with the second law.

The physical origin of this mode switching lies in the idle-level mechanism~\cite{DeOliveira2021}: varying $\eta_0$ changes the idle-level gap without altering the working-level energy difference $A_H-A_C$, thereby redistributing the heat among the four strokes while leaving the fundamental work constraint unchanged.

\begin{figure}[tb]
\centering
\includegraphics[width=0.95\columnwidth]{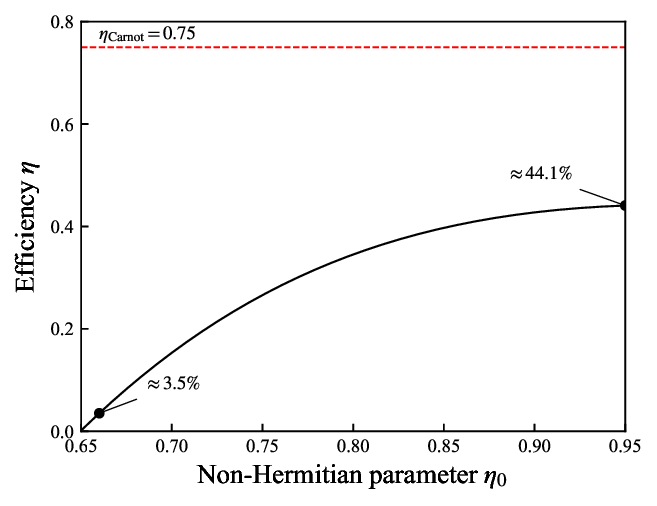}
\caption{Efficiency $\eta$ versus $\eta_0$ (solid line). The horizontal dashed line marks the Carnot limit $\eta_C=0.75$. Parameters are the same as in Fig.~\ref{fig1}.}
\label{fig2}
\end{figure}

Figure~\ref{fig2} shows the engine efficiency as a function of $\eta_0$. Just above the critical point the efficiency is low, but it grows rapidly with increasing $\eta_0$, reaching about $42.7\%$ at $\eta_0=0.9$ and approaching the saturation value $1-A_C/A_H\approx44.15\%$ as $\eta_0\to1^-$. This saturation value coincides with the standard Otto efficiency and remains strictly below the Carnot bound.

The physical origin of this enhancement lies in the asymmetric suppression of the hyperbolic cosine terms at the two temperatures. Because $\beta_c>\beta_h$, the cold-side denominator shrinks faster than the hot-side one as $B$ decreases, so the absorbed heat increases more rapidly than the net work. Since $\eta=W/Q_h$, the efficiency rises accordingly.

\begin{figure}[tb]
\centering
\includegraphics[width=0.85\columnwidth]{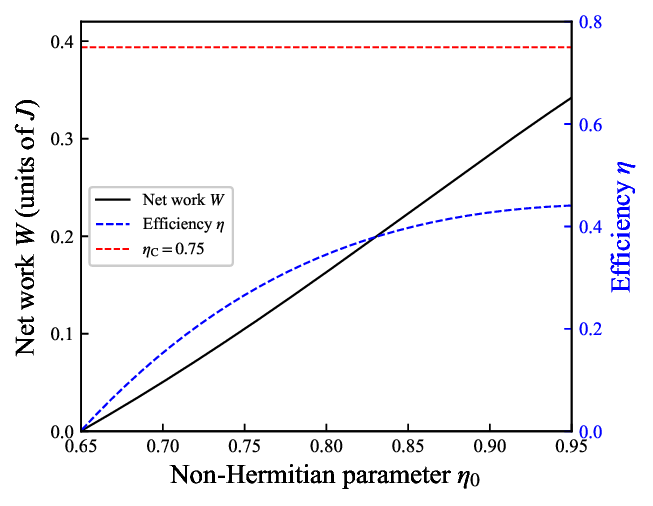}
\caption{Net work $W$ (left axis, solid line, range $0$--$0.42$ in units of $J$) and efficiency $\eta$ (right axis, dashed line, range $0$--$0.8$) versus the non-Hermitian parameter $\eta_0$. The horizontal red dashed line marks the Carnot bound $\eta_C=0.75$. Parameters are the same as in Fig.~\ref{fig1}.}
\label{fig3}
\end{figure}

Figure~\ref{fig3} displays $W$ and $\eta$ on a dual-axis plot. The left vertical axis (net work $W$) spans $0$ to $0.42$, and the right vertical axis (efficiency $\eta$) spans $0$ to $0.8$. In the engine regime both quantities increase monotonically with $\eta_0$, so the non-Hermitian knob enhances work output and efficiency simultaneously. This joint enhancement is a hallmark of the idle-level architecture and cannot be achieved by tuning ordinary working-level parameters, which typically entail a trade-off between power and efficiency.

\begin{figure}[tb]
\centering
\includegraphics[width=0.95\columnwidth]{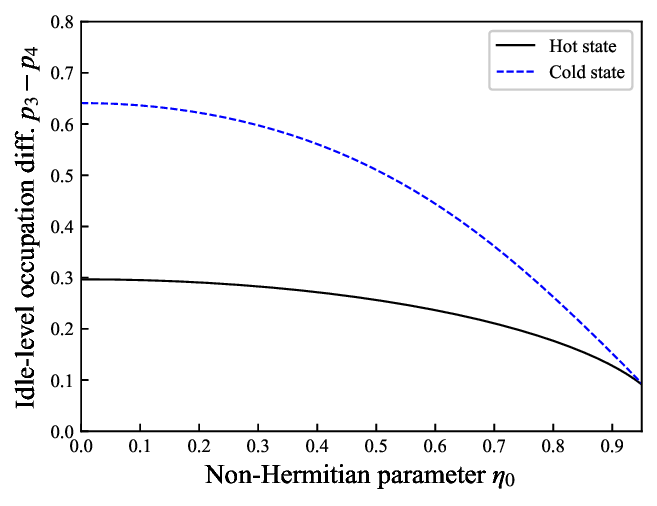}
\caption{Idle-level occupation difference $p_3-p_4$ versus the non-Hermitian parameter $\eta_0$ for the hot state (solid line) and the cold state (dashed line). Parameters are the same as in Fig.~\ref{fig1}.}
\label{fig4}
\end{figure}

To understand the effect at the microscopic level, Fig.~\ref{fig4} plots the idle-level occupation difference $p_3-p_4$ in the hot and cold equilibrium states. At small $\eta_0$ the cold-state occupation difference is substantially larger than the hot-state one. As $\eta_0$ increases, both decrease, but the cold-state value drops much faster. This asymmetric redistribution widens the difference between the hot and cold occupation gaps, which directly drives the absorbed heat $Q_h$. The microscopic origin of the efficiency enhancement is therefore a differential compression of idle-level occupations at the two bath temperatures.

\subsection{Full thermodynamic phase diagram}

To place the results in a broader context, we map the complete operational phase diagram in the $(\eta_0,T_h/T_c)$ plane, following the classification of Ref.~\cite{DeOliveira2021}. Four regimes are identified:
\begin{itemize}
\item Engine: $W>0$, $Q_h>0$ --- converts heat into work.
\item Refrigerator: $W<0$, $Q_h<0$, $Q_c<0$ --- consumes work to pump heat from cold to hot.
\item Accelerator: $W<0$, $Q_h>0$, $Q_c>0$ --- dissipates work and heat into the cold bath.
\item Heater: $W<0$, $Q_h<0$, $Q_c>0$ --- converts work into heat delivered to the cold bath.
\end{itemize}

\begin{figure}[tb]
\centering
\includegraphics[width=0.95\columnwidth]{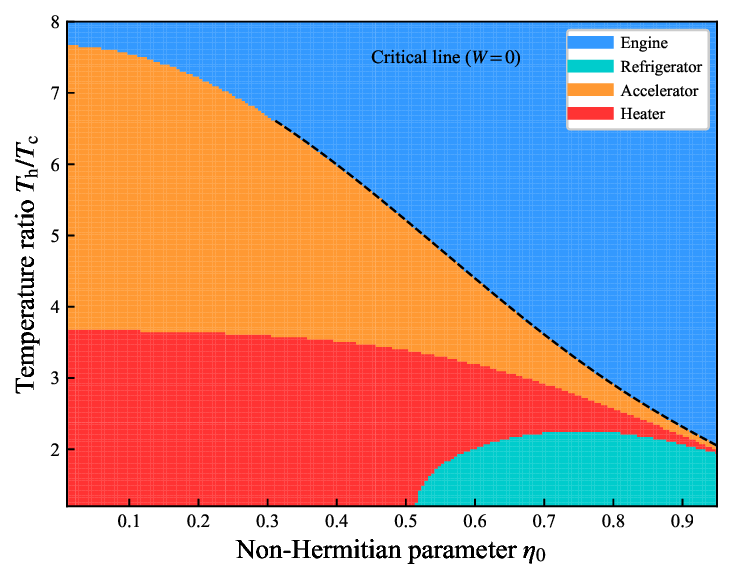}
\caption{Thermodynamic phase diagram in the $(\eta_0,T_h/T_c)$ plane. The horizontal axis is the non-Hermitian parameter $\eta_0$ (range $0.01$--$0.95$); the vertical axis is the temperature ratio $T_h/T_c$ (range $1.2$--$8.0$). The four coloured regions correspond to engine (blue), refrigerator (cyan), accelerator (orange), and heater (red) regimes. The dashed curve marks the $W=0$ critical line. Parameters: $J=1$, $\gamma=0.3$, $h_H=1.5$, $h_C=0.8$, $T_c=0.5$.}
\label{fig5}
\end{figure}

The resulting diagram is shown in Fig.~\ref{fig5}. The horizontal axis is the non-Hermitian parameter $\eta_0$ with range $0.01$ to $0.95$, and the vertical axis is the temperature ratio $T_h/T_c$ with range $1.2$ to $8.0$. The critical line $W=0$, computed by bisection, separates the engine and accelerator regimes and exhibits noticeable curvature. Physically, a larger temperature ratio provides a stronger thermodynamic driving force, so a smaller idle-gap compression (smaller $\eta_0$) suffices to yield positive net work. The engine regime occupies a substantial area for $T_h/T_c>3$, confirming that the heat-engine operation demonstrated above is robust. The existence of all four regimes illustrates the versatility of the non-Hermitian idle-level architecture: the same physical system can perform any of the four fundamental thermodynamic functions simply by adjusting $\eta_0$ and the bath temperatures.

\subsection{Heat current decomposition and reverse heat flow}

To clarify the connection between microscopic occupations and macroscopic efficiency gain, we decompose the absorbed heat into working-level and idle-level contributions,
\begin{equation}
Q_h = Q_{h,w} + Q_{h,i},
\end{equation}
where
\begin{equation}
Q_{h,w}=A_H\bigl[(p_1-p_2)_{T_c}-(p_1-p_2)_{T_h}\bigr]
\label{eq:Qhw}
\end{equation}
is the heat absorbed by the working levels and
\begin{equation}
Q_{h,i}=B\bigl[(p_3-p_4)_{T_c}-(p_3-p_4)_{T_h}\bigr]
\label{eq:Qhi}
\end{equation}
is the heat absorbed by the idle levels. This decomposition is exact and follows directly from the level-resolved internal energy.

\begin{figure}[tb]
\centering
\includegraphics[width=0.95\columnwidth]{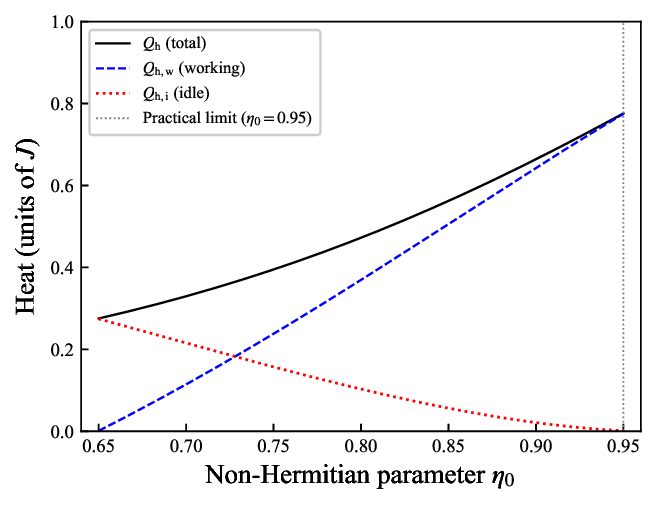}
\caption{Decomposition of the absorbed heat $Q_h$ over the range $\eta_0\in[0.65,0.95]$: total heat $Q_h$ (solid line), working-level contribution $Q_{h,w}$ (dashed line), and idle-level contribution $Q_{h,i}$ (dotted line). The vertical dotted line marks the practical operating limit set by the exceptional point at $\eta_0=0.95$. The idle-level contribution is progressively suppressed toward zero as $\eta_0$ increases, signalling the approach to the reverse heat-flow regime. Parameters are the same as in Fig.~\ref{fig1}.}
\label{fig6}
\end{figure}

Figure~\ref{fig6} plots the three quantities across the engine regime. The working-level contribution dominates the total heat absorption and increases with $\eta_0$, as expected for improved engine operation. The idle-level contribution, by contrast, decreases monotonically and is progressively suppressed toward zero as $\eta_0$ approaches the practical bound of $0.95$.

This suppression is the microscopic manifestation of the reverse heat-flow mechanism identified by de Oliveira and Jonathan~\cite{DeOliveira2021}. As the idle gap shrinks, the net heat absorbed by the idle levels from the hot bath diminishes, and beyond a certain point it reverses sign. When $Q_{h,i}<0$, idle levels transport heat from the cold reservoir back to the hot reservoir, effectively recycling energy that would otherwise be rejected. This internal recycling reduces the net heat that must be supplied by the hot bath for a given work output, which is the thermodynamic origin of the efficiency boost. The occupation redistribution shown in Fig.~\ref{fig4} is therefore not an independent effect but the microscopic cause of the macroscopic reverse heat flow.

Within the physically accessible range $\eta_0<0.95$, $Q_{h,i}$ remains positive but is strongly suppressed, meaning the system approaches the reverse-heat-flow regime without fully entering it. Accessing the fully reversed regime would require operating closer to the exceptional point, where nonadiabatic losses and postselection errors become significant.

\subsection{Comparison with Hermitian idle-level engines}

It is instructive to contrast our non-Hermitian control knob with the two Hermitian idle-level engines previously studied in the literature. We discuss them in turn.

The first is the original model of de Oliveira and Jonathan~\cite{DeOliveira2021}, which is based on an isotropic Heisenberg-type interaction with a single idle level. In that model, the energy spectrum in the notation of Ref.~\cite{DeOliveira2021} reads $-8J$, $-2h$, $0$, $+2h$, where $J$ is the exchange coupling and $h$ is the field. The working levels $\pm 2h$ depend only on $h$ and are unaffected by $J$; the idle level $-8J$ depends only on $J$. Changing $J$ therefore modifies the idle-level gap and shifts the position of the idle level relative to the working levels, but it does not directly change the working-level energies. However, because $-8J$ becomes the ground state for sufficiently large $J$, tuning $J$ strongly affects the ground-state occupation and thus the overall thermodynamic behavior of the cycle. In this sense $J$ controls the idle-level channel, but its influence on the working-level structure is indirect.

The second is the Hermitian XY model studied by Cherubim, de Oliveira, and Jonathan~\cite{Cherubim2022} in the context of nonadiabatic idle-level engines. There the energy spectrum is $\epsilon_{1,4} = \mp\sqrt{4h^2+(J_x-J_y)^2}$ and $\epsilon_{2,3} = \mp(J_x+J_y)$. The working levels $\epsilon_{1,4}$ depend on the field $h$ and on the coupling difference $J_x - J_y$, while the idle levels $\epsilon_{2,3}$ depend on the coupling sum $J_x + J_y$. Consequently, in that model, tuning the exchange couplings inevitably modifies both the idle-level gap and the working-level gap simultaneously.

Our non-Hermitian model is different from both of the above. Here the working-level gap $A_H - A_C$ depends only on $h$ and $\gamma$, while the idle-level gap $B$ depends only on $\eta_0$. The non-Hermitian parameter therefore controls exclusively the idle levels and leaves the working-level difference strictly invariant. This complete decoupling is a distinctive feature of the non-Hermitian realization, and it is not available in either of the Hermitian models. In the Hermitian limit $\eta_0=0$ with our parameters, the system is deeply trapped in the dissipative regime ($W=-0.2672$). No adjustment of the Hermitian parameters can drive it across the critical point into engine operation without simultaneously deforming the working-level structure. The non-Hermitian knob therefore provides a qualitatively new capability: it lifts the system from dissipative behaviour to genuine engine operation while leaving the working substance unchanged.

This strict decoupling is qualitatively different from both Hermitian realizations. In the isotropic Heisenberg model~\cite{DeOliveira2021}, the idle level is itself a product of the exchange coupling $J$; without $J$, the idle level would not exist at all. In our non-Hermitian model, by contrast, the working and idle subspaces are structurally separate from the outset, and $\eta_0$ only tunes the splitting within the already existing idle subspace. This is a fundamentally different control paradigm.

Finally, our model features a symmetric pair of idle levels $\pm B$, whereas the minimal model of Ref.~\cite{DeOliveira2021} has a single idle level. Thermodynamically, the two levels behave as a single effective channel controlled by their occupation difference $p_3-p_4$, so the core idle-level mechanism remains unchanged. However, this two-level structure emerges naturally from the two-qubit XY algebra rather than being imposed by hand, which is what makes the model implementable in standard quantum simulators. In addition, the gap closing at the exceptional point introduces a physical upper bound on the accessible efficiency, a feature absent in the Hermitian idle-level models.

\subsection{Experimental feasibility and limitations}

The two-qubit non-Hermitian XY Hamiltonian can be engineered in trapped-ion quantum simulators using Raman-coupled M\o lmer--S\o rensen schemes combined with optical pumping and ancilla-based postselection. For $^{40}\text{Ca}^+$ setups, coherence times exceeding 100 ms and gate fidelities above 99.9\% are routinely achieved~\cite{Bruzewicz2019}. A complete quasistatic Otto cycle lasting several milliseconds, with $\eta_0$ up to 0.9, is therefore within experimental reach. At these values the predicted efficiency enhancement is already substantial. NMR simulators, which have demonstrated high-fidelity Otto cycles~\cite{Peterson2019}, provide a complementary platform where biorthogonal measurements can be implemented via postselected quantum trajectory schemes.

Several limitations should be kept in mind. First, all results are derived under the quasistatic assumption. Finite cycle times introduce nonadiabatic transitions and quantum friction, especially near the exceptional point where the idle gap shrinks. The effects of nonadiabaticity on idle-level engines have been systematically investigated by Cherubim, de Oliveira, and Jonathan~\cite{Cherubim2022} for the Hermitian XY model. That work showed that qualitative features such as counter-rotating engine operation and non-monotonic efficiency behavior are robust against nonadiabaticity, while quantitative efficiency gains degrade rapidly as the adiabaticity parameter decreases. We expect the same qualitative robustness to hold in our system: the mode-switching and efficiency enhancement phenomena will persist for finite cycle times, even though the magnitude of the efficiency gain will decrease as nonadiabatic mixing between the idle levels increases. In our non-Hermitian model, nonadiabatic transitions are expected to become significant as the idle gap $B$ shrinks near the exceptional point, and the resulting power-efficiency trade-off would be an interesting extension of the present work. Appendix~\ref{app:adiabatic} confirms that the quasistatic regime is accessible for $\eta_0\leq0.95$, but a full finite-time optimization, analogous to Ref.~\cite{Cherubim2022}, remains to be done. Second, the biorthogonal Gibbs state requires continuous postselection, whose success probability decays over successive cycles, reducing the effective work output per physical run. A quantitative accounting of this overhead would require a full quantum trajectory treatment. Both effects are expected to shift the optimal operating point to intermediate values of $\eta_0$.

The robustness of the main findings has been verified by scanning over $\gamma\in[0.1,0.5]$ and $T_h/T_c\in[2,8]$. The critical point shifts by less than 0.05, the monotonic growth of efficiency with $\eta_0$ is preserved in all cases, and the joint enhancement of work and efficiency remains intact.

\section{Summary and outlook}
\label{sec:conclusion}

We have presented a microscopic realization of the idle-level quantum heat engine based on a non-Hermitian two-qubit XY model with a staggered imaginary magnetic field. The spectrum separates naturally into field-dependent working levels and field-decoupled idle levels controlled exclusively by the non-Hermitian parameter. Varying this parameter switches the cycle from a dissipative accelerator regime to a heat engine regime and simultaneously enhances both the net work and the efficiency.

We have mapped the full thermodynamic phase diagram identifying all four fundamental operating regimes, and we have decomposed the heat current into working-level and idle-level contributions. The decomposition reveals that the efficiency enhancement is accompanied by a progressive suppression of the idle-level heat current, which signals the approach to the reverse heat-flow mechanism that underlies the idle-level engine principle. This explicitly connects the microscopic occupation redistribution to the macroscopic thermodynamic phenomenology.

Compared with the original Hermitian idle-level engine, the non-Hermitian control knob offers fully independent tuning of the idle gap without modifying the working levels, and it can lift the system from the dissipative regime into engine operation where Hermitian parameters alone cannot. The symmetric two-idle-level structure emerges naturally from the spin model algebra, making the proposal implementable in standard trapped-ion and NMR simulators.

Several directions merit further investigation. Extending the analysis to the PT-broken phase would reveal whether exceptional-point physics can further modify engine performance. A finite-time optimization including nonadiabatic transitions and postselection costs, along the lines of Ref.~\cite{Cherubim2022}, would clarify the practical power-efficiency trade-off, and would benefit from tools of finite-time quantum thermodynamics and fluctuation theorems~\cite{Campisi2011,Seifert2012,Jarzynski1997,Crooks1999,Esposito2009}. Finally, exploring the interplay between idle-level control and quantum correlations~\cite{Horodecki2009,Chitambar2019}, as well as entanglement~\cite{Wootters1998,Amico2008}, could uncover richer thermodynamic phenomenology. Recent work on negative-temperature quantum heat engines~\cite{Brollo2025} further suggests potential connections between conserved quantities, symmetry, and engine performance in few-level systems.

\section*{Data availability statement}
All data generated or analysed during this study are included in this published article. All numerical and symbolic computations, as well as the generation of all figures, were performed using open-source Python libraries (NumPy, SciPy, Matplotlib) only. No proprietary commercial software was used. The complete Python source code is available from the corresponding author upon reasonable request.

\begin{acknowledgments}
This work was supported by the Natural Science Foundation of Xinjiang Uygur Autonomous Region (Grants Nos. 2023D01A42 and 2023D01B50), the National Natural Science Foundation of China (Grant No. 12405025), and the Doctoral Research Startup Foundation of Xinjiang Normal University (Grants Nos. XINUZBS2433 and XJNUZBS2414).
\end{acknowledgments}

\appendix

\section{Diagonalization and biorthonormal basis}
\label{app:diag}

The Hamiltonian decomposes as $H=H_{\rm work}\oplus H_{\rm idle}$ with
\begin{equation}
H_{\rm work}=\begin{pmatrix}-2Jh_0 & -2J\gamma\\ -2J\gamma & 2Jh_0\end{pmatrix},
\qquad
H_{\rm idle}=\begin{pmatrix}-2iJ\eta_0 & -2J\\ -2J & 2iJ\eta_0\end{pmatrix}.
\label{eq:Hblocks}
\end{equation}

The working-level eigenvalues are $E_{1,2}=\mp2J\sqrt{h_0^2+\gamma^2}$. Defining
$d_1=h_0+\sqrt{h_0^2+\gamma^2}$ and $d_2=h_0-\sqrt{h_0^2+\gamma^2}$, the
normalized right eigenvectors are
\begin{equation}
|\Psi_1^R\rangle=\frac{1}{\sqrt{\gamma^2+d_1^2}}
\bigl(d_1|\uparrow\uparrow\rangle+\gamma|\downarrow\downarrow\rangle\bigr),
\quad
|\Psi_2^R\rangle=\frac{1}{\sqrt{\gamma^2+d_2^2}}
\bigl(d_2|\uparrow\uparrow\rangle+\gamma|\downarrow\downarrow\rangle\bigr).
\label{eq:psiwork}
\end{equation}
Since $H_{\rm work}$ is Hermitian, the left eigenvectors satisfy $\langle\Psi_k^L|=|\Psi_k^R\rangle^\dagger$.

The idle-level eigenvalues are $E_{3,4}=\mp2J\sqrt{1-\eta_0^2}$. Defining
$d_3=i\eta_0+\sqrt{1-\eta_0^2}$, $d_4=i\eta_0-\sqrt{1-\eta_0^2}$,
$d_5=-i\eta_0+\sqrt{1-\eta_0^2}$, $d_6=-i\eta_0-\sqrt{1-\eta_0^2}$, the
eigenvectors are
\begin{equation}
|\Psi_{3,4}^R\rangle=\frac{1}{\sqrt{1+|d_{3,4}|^2}}
\bigl(d_{3,4}|\uparrow\downarrow\rangle+|\downarrow\uparrow\rangle\bigr),
\quad
|\Psi_{3,4}^L\rangle=\frac{1}{\sqrt{1+|d_{5,6}|^2}}
\bigl(d_{5,6}|\uparrow\downarrow\rangle+|\downarrow\uparrow\rangle\bigr).
\label{eq:psiidle}
\end{equation}
After rescaling by the normalization constants
$c_{3,4}=2\sqrt{1-\eta_0^2}(\sqrt{1-\eta_0^2}\pm i\eta_0)$, the basis satisfies
biorthonormality $\langle\Psi_i^L|\Psi_j^R\rangle=\delta_{ij}$ and completeness
$\sum_i|\Psi_i^R\rangle\langle\Psi_i^L|=\mathbb{I}$, both verified symbolically.

\section{Thermodynamic quantities}
\label{app:thermal}

The internal energy can be computed both from the spectral decomposition and from the partition function, yielding identical results. The level occupations are
$p_1=e^{\beta A}/Z$, $p_2=e^{-\beta A}/Z$, $p_3=e^{\beta B}/Z$, $p_4=e^{-\beta B}/Z$,
with $Z=2\cosh(\beta A)+2\cosh(\beta B)$. Table~\ref{tab1} compares the analytical
internal energies with exact numerical diagonalization for representative values of
$\eta_0$, confirming agreement to machine precision. The table also verifies the
first law $W=Q_h-Q_c$ across the full parameter range.

\begin{table}[tb]
\caption{Analytical and numerical internal energies and first-law verification.
Parameters: $J=1$, $\gamma=0.3$, $h_H=1.5$, $h_C=0.8$, $T_h=2.0$, $T_c=0.5$.}
\label{tab1}
\centering
\begin{tabular}{cccccc}
\toprule
$\eta_0$ & $(h,T)$ & $U_{\rm ana}$ & $U_{\rm num}$ & $|\Delta U|$ & $|W-(Q_h-Q_c)|$\\
\midrule
0.0 & $(h_H,T_h)$ & $-2.2934$ & $-2.2934$ & $<10^{-14}$ & $<10^{-14}$\\
0.5 & $(h_H,T_h)$ & $-2.2081$ & $-2.2081$ & $<10^{-14}$ & $<10^{-14}$\\
0.9 & $(h_H,T_h)$ & $-2.0275$ & $-2.0275$ & $<10^{-14}$ & $<10^{-14}$\\
0.0 & $(h_C,T_c)$ & $-1.8934$ & $-1.8934$ & $<10^{-14}$ & $<10^{-14}$\\
0.5 & $(h_C,T_c)$ & $-1.7172$ & $-1.7172$ & $<10^{-14}$ & $<10^{-14}$\\
0.9 & $(h_C,T_c)$ & $-1.5619$ & $-1.5619$ & $<10^{-14}$ & $<10^{-14}$\\
\bottomrule
\end{tabular}
\end{table}

\section{Otto cycle quantities}
\label{app:workheat}

With $A_H=2J\sqrt{h_H^2+\gamma^2}$, $A_C=2J\sqrt{h_C^2+\gamma^2}$,
$B=2J\sqrt{1-\eta_0^2}$, $\beta_h=1/T_h$, $\beta_c=1/T_c$, the internal energies at
the four cycle points are
\begin{align}
U_1 &= -\frac{A_H\sinh(\beta_h A_H)+B\sinh(\beta_h B)}
{\cosh(\beta_h A_H)+\cosh(\beta_h B)}, \\
U_2 &= -\frac{A_C\sinh(\beta_h A_H)+B\sinh(\beta_h B)}
{\cosh(\beta_h A_H)+\cosh(\beta_h B)}, \\
U_3 &= -\frac{A_C\sinh(\beta_c A_C)+B\sinh(\beta_c B)}
{\cosh(\beta_c A_C)+\cosh(\beta_c B)}, \\
U_4 &= -\frac{A_H\sinh(\beta_c A_C)+B\sinh(\beta_c B)}
{\cosh(\beta_c A_C)+\cosh(\beta_c B)}.
\label{eq:U1234}
\end{align}

Forming $Q_h=U_1-U_4$ and $Q_c=U_2-U_3$, one finds that the $B$-proportional
numerator terms cancel exactly in the net work $W=Q_h-Q_c$, yielding
Eq.~\eqref{eq:W} of the main text. This cancellation is the algebraic reason why
idle levels modify the absorbed heat more strongly than the net work.

In the Hermitian limit $\eta_0=0$, $B=2J$, and all expressions reduce to those of
an ordinary four-level Hermitian Otto engine, confirming that the non-Hermitian
framework recovers standard thermodynamics smoothly.

\section{Adiabatic condition and finite-time effects}
\label{app:adiabatic}

During the adiabatic strokes only the field $h$ is varied while $\eta_0$ is held
fixed. The idle-level eigenvalues depend only on $\eta_0$, so the idle gap remains
constant throughout these strokes and
$\langle\Psi_3^L|\dot H|\Psi_4^R\rangle=0$; the adiabatic condition for the
idle-level pair is therefore trivially satisfied.

For the working-level pair, the instantaneous gap is $2A(h)$, and the coupling
matrix element is
\begin{equation}
\langle\Psi_1^L|\dot H|\Psi_2^R\rangle
= \frac{2J\gamma\dot h}{h^2+\gamma^2}.
\end{equation}
The adiabatic condition reads
\begin{equation}
\left|\frac{\langle\Psi_1^L|\dot H|\Psi_2^R\rangle}{(E_1-E_2)^2}\right| \ll 1.
\end{equation}
For a linear field ramp from $h_C$ to $h_H$ over stroke time $\tau/2$, the
left-hand side peaks near $h=h_C$ and scales as
$\epsilon_{\rm ad} \sim J\gamma(h_H-h_C)/(2\tau A_C^3)$.
For our parameters, $\epsilon_{\rm ad}\lesssim 10^{-3}$ for $\tau\gtrsim 100/J$,
which is well within the coherence times of state-of-the-art trapped-ion
simulators.

The restriction $\eta_0<0.95$ is imposed not because of the field sweep but to
stay away from the exceptional point at $\eta_0=1$, where the biorthogonal
eigenbasis becomes ill-conditioned. Near the exceptional point, even small
perturbations can mix the idle eigenstates strongly, leading to enhanced
nonadiabatic transitions and quantum friction that degrade both power and
efficiency. A full finite-time optimization including this effect is beyond the
scope of the present work.


\begin{thebibliography}{99}

\bibitem{Alicki1979} Alicki R 1979 \href{https://doi.org/10.1088/0305-4470/12/5/007}{\textit{J. Phys. A: Math. Gen.} \textbf{12}, L103}

\bibitem{Kosloff1984} Kosloff R 1984 \href{https://doi.org/10.1063/1.446860}{\textit{J. Chem. Phys.} \textbf{80}, 1625}

\bibitem{Geva1992} Geva E and Kosloff R 1992 \href{https://doi.org/10.1063/1.461951}{\textit{J. Chem. Phys.} \textbf{96}, 3054}

\bibitem{Abah2012} Abah O, Ro\ss nagel J, Jacob G, Deffner S, Schmidt-Kaler F, Singer K and Lutz E 2012 \href{https://doi.org/10.1103/PhysRevLett.109.203006}{\textit{Phys. Rev. Lett.} \textbf{109}, 203006}

\bibitem{Rossnagel2016} Ro\ss nagel J, Dawkins S T, Tolazzi K N, Abah O, Lutz E, Schmidt-Kaler F and Singer K 2016 \href{https://doi.org/10.1126/science.aad6326}{\textit{Science} \textbf{352}, 325}

\bibitem{Klatzow2019} Klatzow J, Becker J N, Ledingham P M, Weinzettl C, Kaczmarek K T, Saunders D J, Nunn J, Walmsley I A, Uzdin R and Poem E 2019 \href{https://doi.org/10.1103/PhysRevLett.122.110601}{\textit{Phys. Rev. Lett.} \textbf{122}, 110601}

\bibitem{Peterson2019} Peterson J P S, Batalh\~ao T B, Herrera M, Souza A M, Sarthour R S, Oliveira I S and Serra R M 2019 \href{https://doi.org/10.1103/PhysRevLett.123.240601}{\textit{Phys. Rev. Lett.} \textbf{123}, 240601}

\bibitem{Quan2007} Quan H T, Liu Y, Sun C P and Nori F 2007 \href{https://doi.org/10.1103/PhysRevE.76.031105}{\textit{Phys. Rev. E} \textbf{76}, 031105}

\bibitem{Kosloff2017} Kosloff R and Rezek Y 2017 \href{https://doi.org/10.3390/e19040136}{\textit{Entropy} \textbf{19}, 136}

\bibitem{Gemmer2009} Gemmer J, Michel M and Mahler G 2009 \href{https://doi.org/10.1007/978-3-540-70510-9}{\textit{Quantum Thermodynamics}} (Berlin: Springer)

\bibitem{Vinjanampathy2016} Vinjanampathy S and Anders J 2016 \href{https://doi.org/10.1080/00107514.2016.1201896}{\textit{Contemp. Phys.} \textbf{57}, 545}

\bibitem{Goold2016} Goold J, Huber M, Riera A, del Rio L and Skrzypczyk P 2016 \href{https://doi.org/10.1088/1751-8113/49/14/143001}{\textit{J. Phys. A: Math. Theor.} \textbf{49}, 143001}

\bibitem{Binder2018} Binder F, Correa L A, Gogolin C, Anders J and Adesso G (eds) 2018 \href{https://doi.org/10.1007/978-3-319-99046-0}{\textit{Thermodynamics in the Quantum Regime}} (Cham: Springer)

\bibitem{Bender1998} Bender C M and Boettcher S 1998 \href{https://doi.org/10.1103/PhysRevLett.80.5243}{\textit{Phys. Rev. Lett.} \textbf{80}, 5243}

\bibitem{Bender2007} Bender C M 2007 \href{https://doi.org/10.1088/0034-4885/70/6/R03}{\textit{Rep. Prog. Phys.} \textbf{70}, 947}

\bibitem{Feng2017} Feng L, El-Ganainy R and Ge L 2017 \href{https://doi.org/10.1038/nphoton.2017.21}{\textit{Nat. Photon.} \textbf{11}, 752}

\bibitem{Ozdemir2019} \"Ozdemir \c{S} K, Rotter S, Nori F and Yang L 2019 \href{https://doi.org/10.1038/s41563-019-0304-9}{\textit{Nat. Mater.} \textbf{18}, 783}

\bibitem{Ruter2010} R\"uter C E, Makris K G, El-Ganainy R, Christodoulides D N, Segev M and Kip D 2010 \href{https://doi.org/10.1038/nphys1515}{\textit{Nat. Phys.} \textbf{6}, 192}

\bibitem{Konotop2016} Konotop V V, Yang J and Zezyulin D A 2016 \href{https://doi.org/10.1103/RevModPhys.88.035002}{\textit{Rev. Mod. Phys.} \textbf{88}, 035002}

\bibitem{ElGanainy2018} El-Ganainy R, Makris K G, Khajavikhan M, Musslimani Z H, Rotter S and Christodoulides D N 2018 \href{https://doi.org/10.1038/nphys4323}{\textit{Nat. Phys.} \textbf{14}, 11}

\bibitem{Ashida2020} Ashida Y, Gong Z and Ueda M 2020 \href{https://doi.org/10.1080/00018732.2021.1883316}{\textit{Adv. Phys.} \textbf{69}, 249}

\bibitem{Lin2016} Lin S and Song Z 2016 \href{https://doi.org/10.1088/1751-8113/49/47/475301}{\textit{J. Phys. A: Math. Theor.} \textbf{49}, 475301}

\bibitem{Insinga2018} Insinga A, Andresen B, Salamon P and Kosloff R 2018 \href{https://doi.org/10.1103/PhysRevE.97.062153}{\textit{Phys. Rev. E} \textbf{97}, 062153}

\bibitem{Zhang2022} Zhang J-W, Zhang J-Q, Ding G-Y, Li J-C, Bu J-T, Wang B, Yan L-L, Su S-L, Chen L, Nori F, \"Ozdemir \c{S} K, Zhou F, Jing H and Feng M 2022 \href{https://doi.org/10.1038/s41467-022-33932-9}{\textit{Nat. Commun.} \textbf{13}, 6225}

\bibitem{Zhou2021} Zhou Z-Y, Chen Y and Li Y 2021 \href{https://doi.org/10.1103/PhysRevE.104.034107}{\textit{Phys. Rev. E} \textbf{104}, 034107}

\bibitem{Khandelwal2021} Khandelwal S, Brunner N and Haack G 2021 \href{https://doi.org/10.1103/PRXQuantum.2.040346}{\textit{PRX Quantum} \textbf{2}, 040346}

\bibitem{DeOliveira2021} de Oliveira T R and Jonathan D 2021 \href{https://doi.org/10.1103/PhysRevE.104.044133}{\textit{Phys. Rev. E} \textbf{104}, 044133}

\bibitem{DeAlmeida2024} Anka M F, de Oliveira T R and Jonathan D 2024 \href{https://doi.org/10.1103/PhysRevE.109.064129}{\textit{Phys. Rev. E} \textbf{109}, 064129}

\bibitem{Cherubim2022} Cherubim C, de Oliveira T R and Jonathan D 2022 \href{https://doi.org/10.1103/PhysRevE.105.044120}{\textit{Phys. Rev. E} \textbf{105}, 044120}

\bibitem{Li2023} Li Y, Zhang P-P, Hu L-Z, Xu Y-L and Kong X-M 2023 \href{https://doi.org/10.1007/s11128-023-04031-z}{\textit{Quantum Inf. Process.} \textbf{22}, 277}

\bibitem{Gardas2016} Gardas B, Deffner S and Saxena A 2016 \href{https://doi.org/10.1038/srep23408}{\textit{Sci. Rep.} \textbf{6}, 23408}

\bibitem{Mostafazadeh2010} Mostafazadeh A 2010 \href{https://doi.org/10.1142/S0219887810001833}{\textit{Int. J. Geom. Methods Mod. Phys.} \textbf{7}, 1191}

\bibitem{Bruzewicz2019} Bruzewicz C D, Chiaverini J, McConnell R and Sage J M 2019 \href{https://doi.org/10.1063/1.5088164}{\textit{Appl. Phys. Rev.} \textbf{6}, 021314}

\bibitem{Campisi2011} Campisi M, H\"anggi P and Talkner P 2011 \href{https://doi.org/10.1103/RevModPhys.83.771}{\textit{Rev. Mod. Phys.} \textbf{83}, 771}

\bibitem{Seifert2012} Seifert U 2012 \href{https://doi.org/10.1088/0034-4885/75/12/126001}{\textit{Rep. Prog. Phys.} \textbf{75}, 126001}

\bibitem{Jarzynski1997} Jarzynski C 1997 \href{https://doi.org/10.1103/PhysRevLett.78.2690}{\textit{Phys. Rev. Lett.} \textbf{78}, 2690}

\bibitem{Crooks1999} Crooks G E 1999 \href{https://doi.org/10.1023/A:1023066317855}{\textit{J. Stat. Phys.} \textbf{90}, 1481}

\bibitem{Esposito2009} Esposito M, Harbola U and Mukamel S 2009 \href{https://doi.org/10.1103/RevModPhys.81.1665}{\textit{Rev. Mod. Phys.} \textbf{81}, 1665}

\bibitem{Horodecki2009} Horodecki R, Horodecki P, Horodecki M and Horodecki K 2009 \href{https://doi.org/10.1103/RevModPhys.81.865}{\textit{Rev. Mod. Phys.} \textbf{81}, 865}

\bibitem{Chitambar2019} Chitambar E and Gour G 2019 \href{https://doi.org/10.1103/RevModPhys.91.025001}{\textit{Rev. Mod. Phys.} \textbf{91}, 025001}

\bibitem{Wootters1998} Wootters W K 1998 \href{https://doi.org/10.1103/PhysRevLett.80.2245}{\textit{Phys. Rev. Lett.} \textbf{80}, 2245}

\bibitem{Amico2008} Amico L, Fazio R, Osterloh A and Vedral V 2008 \href{https://doi.org/10.1103/RevModPhys.80.517}{\textit{Rev. Mod. Phys.} \textbf{80}, 517}

\bibitem{Brollo2025} Brollo A, del Campo A and Bastianello A 2025 \href{https://doi.org/10.1038/s41467-025-10593-x}{\textit{Nat. Commun.} \textbf{16}, 10593}

\end{thebibliography}
\end{document}